\documentclass[11pt]{article}
\usepackage{moriond,epsfig}

\def\PLB{{\em Phys. Lett.}  B}

\def\COMP{{\em Computer Physics Commun}}

\def\be{\begin{equation}}
\def\ee{\end{equation}}
\def\bea{\begin{eqnarray}}
\def\eea{\end{eqnarray}}

\RequirePackage{xspace}

\usepackage{relsize}
\def\babar{\mbox{\slshape B\kern-0.1em{\smaller A}\kern-0.1em
    B\kern-0.1em{\smaller A\kern-0.2em R}}}

\def\d     {\ensuremath{d}\xspace}

\def\s     {\ensuremath{s}\xspace}

\def\b     {\ensuremath{b}\xspace}
\def\bbar  {\ensuremath{\overline b}\xspace}
\def\bbbar {\ensuremath{b\overline b}\xspace}

\def\pip   {\ensuremath{\pi^+}\xspace}
\def\pim   {\ensuremath{\pi^-}\xspace}

\def\Kbar  {\kern 0.2em\overline{\kern -0.2em K}{}\xspace}

\def\Kz    {\ensuremath{K^0}\xspace}
\def\Kzb   {\ensuremath{\Kbar^0}\xspace}
\def\KzKzb {\ensuremath{\Kz \kern -0.16em \Kzb}\xspace}
\def\Kp    {\ensuremath{K^+}\xspace}
\def\Km    {\ensuremath{K^-}\xspace}
\def\Kpm   {\ensuremath{K^\pm}\xspace}

\def\KpKm  {\ensuremath{\Kp \kern -0.16em \Km}\xspace}
\def\KS    {\ensuremath{K^0_{\scriptscriptstyle S}}\xspace}

\def\Kstarz  {\ensuremath{K^{*0}}\xspace}

\def\Kstar   {\ensuremath{K^*}\xspace}

\def\Dbar    {\kern 0.2em\overline{\kern -0.2em D}{}\xspace}

\def\Dz      {\ensuremath{D^0}\xspace}
\def\Dzb     {\ensuremath{\Dbar^0}\xspace}
\def\DzDzb   {\ensuremath{\Dz {\kern -0.16em \Dzb}}\xspace}
\def\Dp      {\ensuremath{D^+}\xspace}
\def\Dm      {\ensuremath{D^-}\xspace}

\def\DpDm    {\ensuremath{\Dp {\kern -0.16em \Dm}}\xspace}
\def\Dstar   {\ensuremath{D^*}\xspace}

\def\B       {\ensuremath{B}\xspace}

\def\Bbar    {\kern 0.18em\overline{\kern -0.18em B}{}\xspace}
\def\Nbar    {\kern 0.18em\overline{\kern -0.18em N}{}\xspace}

\def\Bz      {\ensuremath{B^0}\xspace}
\def\Bzb     {\ensuremath{\Bbar^0}\xspace}
\def\BzBzb   {\ensuremath{\Bz {\kern -0.16em \Bzb}}\xspace}
\def\myBzBzb {\ensuremath{\Bz {\kern -0.16em - \Bzb}}\xspace}
\def\Bu      {\ensuremath{B^+}\xspace}
\def\Bub     {\ensuremath{B^-}\xspace}

\def\BpBm    {\ensuremath{\Bu {\kern -0.16em \Bub}}\xspace}
\def\Bs      {\ensuremath{B^0_s}\xspace}
\def\Bd      {\ensuremath{B^0_d}\xspace}
\def\Bc      {\ensuremath{B_c}\xspace}
\def\Bsb     {\ensuremath{\Bbar^0_s}\xspace}
\def\BsBsb   {\ensuremath{\Bs {\kern -0.16em-\Bsb}}\xspace}

\def\BorBbar    {\kern 0.18em\optbar{\kern -0.18em B}{}\xspace}
\def\DorDbar    {\kern 0.18em\optbar{\kern -0.18em D}{}\xspace}
\def\KorKbar    {\kern 0.18em\optbar{\kern -0.18em K}{}\xspace}

\def\jpsi     {\ensuremath{{J\mskip -3mu/\mskip -2mu\psi\mskip 2mu}}\xspace}

\mathchardef\Upsilon="7107
\def\Y#1S{\ensuremath{\Upsilon{(#1S)}}\xspace}

\mathchardef\Deltares="7101
\mathchardef\Xi="7104
\mathchardef\Lambda="7103
\mathchardef\Sigma="7106
\mathchardef\Omega="710A

\def\Deltabar{\kern 0.25em\overline{\kern -0.25em \Deltares}{}\xspace}
\def\Lbar{\kern 0.2em\overline{\kern -0.2em\Lambda\kern 0.05em}\kern-0.05em{}\xspace}
\def\Sigbar{\kern 0.2em\overline{\kern -0.2em \Sigma}{}\xspace}
\def\Xibar{\kern 0.2em\overline{\kern -0.2em \Xi}{}\xspace}
\def\Obar{\kern 0.2em\overline{\kern -0.2em \Omega}{}\xspace}
\def\Nbar{\kern 0.2em\overline{\kern -0.2em N}{}\xspace}
\def\Xb{\kern 0.2em\overline{\kern -0.2em X}{}\xspace}

\def\bpsiks     {\ensuremath{\Bz \to \jpsi \KS}\xspace}

\def\bphiks     {\ensuremath{\Bz \to \mphi \KS}\xspace}

\def\bskpkm     {\ensuremath{\Bz_{s} \to \Kp \Km}\xspace}

\def\bropi      {\ensuremath{\Bz \to \rho \pi}\xspace}

\def\bdpippim       {\ensuremath{\Bz_d \to \pip \pim}\xspace}
\def\bskpkm         {\ensuremath{\Bz_s \to \Kp \Km}\xspace}

\def\bsdsp          {\ensuremath{\Bz_s \to D_s^- \pip}\xspace}

\def\bsdsmpkpm      {\ensuremath{\Bz_s \to D_s^{\mp} \Kpm}\xspace}
\def\bsjpsiphi      {\ensuremath{\Bz_s \to \jpsi \mphi}\xspace}

\def\bzdzkst     {\ensuremath{\Bz \to \Dz \Kstar}\xspace}
\def\bzdzbkstz   {\ensuremath{\Bz \to \Dzb \Kstarz}\xspace}
\def\bzdzcpkstz  {\ensuremath{\Bz \to \Dz_{CP} \Kstarz}\xspace}

\def\bzsmumu          {\ensuremath{\Bz_s \to \mu^+\mu^-}\xspace}
\def\bzdkstargamma    {\ensuremath{\Bz_d \to \Kstarz \gamma}\xspace}
\def\bzsphigamma      {\ensuremath{\Bz_s \to \mphi \gamma}\xspace}
\def\bzsphiphi        {\ensuremath{\Bz_s \to \mphi \mphi}\xspace}
\def\bzdmumukst       {\ensuremath{\Bz_d \to \mu^+\mu^- \Kstarz}\xspace}

\def\mphi       {\mbox{$\phi$}\xspace}

\newcommand{\tev}{\ensuremath{\mathrm{\,Te\kern -0.1em V}}\xspace}
\newcommand{\gev}{\ensuremath{\mathrm{\,Ge\kern -0.1em V}}\xspace}
\newcommand{\mev}{\ensuremath{\mathrm{\,Me\kern -0.1em V}}\xspace}
\newcommand{\kev}{\ensuremath{\mathrm{\,ke\kern -0.1em V}}\xspace}
\newcommand{\ev}{\ensuremath{\mathrm{\,e\kern -0.1em V}}\xspace}
\newcommand{\gevc}{\ensuremath{{\mathrm{\,Ge\kern -0.1em V\!/}c}}\xspace}
\newcommand{\mevc}{\ensuremath{{\mathrm{\,Me\kern -0.1em V\!/}c}}\xspace}
\newcommand{\gevcc}{\ensuremath{{\mathrm{\,Ge\kern -0.1em V\!/}c^2}}\xspace}
\newcommand{\mevcc}{\ensuremath{{\mathrm{\,Me\kern -0.1em V\!/}c^2}}\xspace}

\def\mus  {\ensuremath{\rm \,\mus}\xspace}

\def\mus        {\ensuremath{\,\mu{\rm s}}\xspace}    

\def\to                 {\ensuremath{\rightarrow}\xspace}

\def\pep2{PEP-II}

\def\gsim{{~\raise.15em\hbox{$>$}\kern-.85em
          \lower.35em\hbox{$\sim$}~}\xspace}
\def\lsim{{~\raise.15em\hbox{$<$}\kern-.85em
          \lower.35em\hbox{$\sim$}~}\xspace}

\def\CP                {\ensuremath{C\!P}\xspace}

\def\deltams{\ensuremath{{\rm \Delta}m_s}\xspace}

\xspace

\def\jetset74   {\mbox{\tt Jetset \hspace{-0.5em}7.\hspace{-0.2em}4}\xspace}

\begin{document}
\vspace*{4cm}
\title{OVERVIEW OF LHCb}
\author{HERV\'E TERRIER (on behalf of the LHCb collaboration)}
\address{Laboratoire d'Annecy-le-Vieux de Physique des Particules, 
  9 Chemin de Bellevue, 
  BP 110 F, 
  74941 Annecy-le-Vieux CEDEX}
\maketitle\abstracts{An overview of LHCb experiment is given, 
  focusing on detector, trigger and expected physics performances.
  LHCb is a second generation b physics experiment design to do precise 
  measurements of CP violation in B meson system and to study b hadron rare 
  decays.}
\section{Introduction}
\begin{figure}[h]
  \begin{center}
    \epsfig{figure=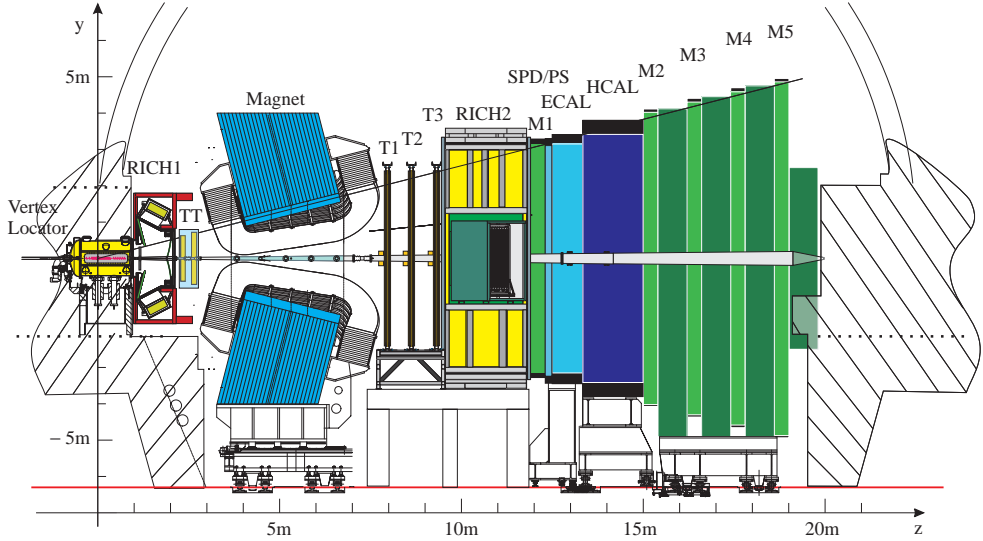,height=2.8in}
    \caption{Side view of LHCb detector.\label{fig:detector}}
  \end{center}
\end{figure}

The LHCb experiment \cite{lhcb_tp,lhcb_tdr} has been designed to study CP violation 
and rare phenomena in \B meson decays with very high precision. The physics goal 
being to provide an understanding of quark flavor physics, and possibly to reveal 
physics beyond the Standard Model. The experiment will be based at CERN and will 
study the decay of \b quarks from \bbbar pairs produced in proton-proton 
collision. It will operate at an average luminosity of 
$2\times10^{32}$cm$^{-2}$s$^{-1}$, much lower than the maximum design luminosity of the 
LHC, in order to have a number of interactions per crossing dominated by single 
interaction. This facilitates the triggering and reconstruction by assuring low 
channel occupancy. In addition, the radiation damages are more manageable. 
To achieve its physics goals, the experiment will have to perform,  
a high track reconstruction efficiency, a good $\pi-K$ separation for 
momenta from few to $\sim$100 GeV/c, a very good propertime resolution 
($\sim$40 fs), and, a high trigger efficiency (both for final states including 
leptons and for those with hadrons only). These needs have led to the detector 
design that is represented on Fig. \ref{fig:detector} and that is described in the 
next section.
\section{LHCb detector description}
\begin{figure}[h]
  \begin{center}
    \epsfig{figure=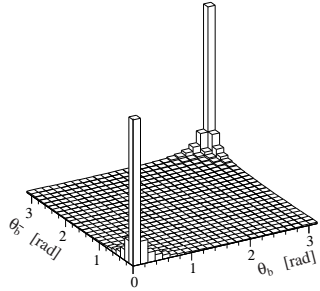,height=1.5in}
    \caption{Polar angles of the \b and \bbar hadrons produced in proton-proton
      collisions with $\sqrt s = 14$ TeV calculated by the PYTHIA event generator.}
    \label{fig:bbbarangle}
  \end{center}
\end{figure}

LHCb detector is a single-arm spectrometer with a forward angular
coverage from 10 mrad to 300 (250) mrad in the bending (non-bending) plane. The 
choice of the geometry has been motivated by the fact that at high energies both
the \b and \bbar hadrons are produced at small angles with respect to the beam 
pipe \cite{lhcb_tp}. This is illustrated by Fig. \ref{fig:bbbarangle} obtained
from PYTHIA event generator \cite{pythia}.
Figure \ref{fig:detector} shows a side view of the detector that consists of the 
beam pipe, the vertex detector (VELO), the dipole magnet, the tracking system 
(TT, T1-T3), two Ring Imaging Cherenkov detectors (RICH1 and RICH2), 
the calorimeter system (SPD/PS, ECAL and HCAL) and the muon system (M1-M5). 
The VELO has to provide precise measurements of track 
coordinates close to the interaction region in order to get good proper time
resolution. It is made of circular silicon stations placed along and 
perpendicularly to the beam axis. With the excellent momentum resolution achieved
by the tracking system, the proper time of reconstructed \B mesons can be measured with a 
resolution of $\sim$40 fs \cite{lhcb_tdr}.
The tracking system is composed of the dipole magnet which is a 
warm magnet and has a field integral of 4 Tm, a Trigger Tracker (TT) located
in front of the magnet entrance and three tracking station (T1-T3) placed 
behind the magnet.
TT stations are made of silicon sensors and play two roles. Firstly, it will be used 
in the Level-1 trigger to assign transverse momentum information to large impact 
parameter tracks. Secondly, it will be used in the offline analysis to 
reconstruct tracks, in particular the decay products of long-lived neutral particle 
that decay outside the VELO.
Each T station consists of an Inner Tracker (IT) close to the beam 
pipe and an Outer Tracker (OT) surrounding the IT. The IT is made 
of silicon strip detectors and the OT of straw tubes.
Charged tracks are reconstructed with a high efficiency of $\sim$95\% with a low rate
of wrongly reconstructed tracks, which does not introduce significant additionnal 
combinatorial background in the reconstructed \B meson signals.
RICH system is composed of two elements located in front of TT 
and behind T stations. To cover the momentum range from few to $\sim$100 GeV/c, 
three different radiators have been chosen : silica aerogel and two fluorocarbon 
gases, C$_4$F$_{10}$ and CF$_4$. In this momentum range, the average efficiency
for kaon identification is 88\% for an average pion misidentification rate of 3\%.
The main purpose of the calorimeter system is to identify electrons
and hadrons and to provide measurements of their energy and position. These 
measurements are used by the trigger system and for the off-line analysis. 
The structure consists of four elements, a scintillator pad detector (SPD), a 
preshower (PS), an electromagnetic calorimeter (ECAL) and an hadronic calorimeter 
(HCAL). These elements employ similar technologies, i.e. scintillators coupled to 
wavelength-shifting fibers read out by fast photodectors. The electron 
identification efficiency for tracks in ECAL acceptance is $\sim$94\% with a pion
misidentification rate of $\sim$0.7\%.
The muon detector is used to identify muons for the trigger and 
off-line analysis. It consists of five stations, M1 in front of the 
calorimeter system and M2-M5 behind the calorimeter, interleaved with iron 
shielding plates. Each station is made of four layers of Multi Wire Proportional
Chambers (MWPC) except for M1 which is made of two. For tracks in muon 
detector acceptance, muon identification efficiency is $\sim$93\% with a pion
misidentification rate of $\sim$1\%.
\section{LHCb trigger}
The trigger \cite{triggertdr} is one of the biggest challenge for the LHCb 
experiment, the \bbbar pair creation cross section being less than 1\% of the total
cross section. It is designed to distinguish events containing \B mesons from 
minimum-bias events through the presence of particles with a large transverse 
momentum ($p_T$) and the existence of secondary vertices. The trigger is divided 
in three \mbox{levels :} Level-0 (L0) implemented in custom electronics, Level-1 
(L1) and High Level Trigger (HLT) both executed in farm processors. The L0 trigger
will reduce the 40 MHz LHC beam crossing rate to 1 MHz. The events are triggered by 
requiring at least one lepton or hadron with a $p_T$ exceeding 1 to 3 GeV/c. Events 
can also be rejected based on global event variables such as track multiplicities 
and number of interactions. 
The L1 trigger selects events with an output rate of 40 kHz. The L1 algorithm 
reconstructs tracks in the VELO and matches these tracks to L0 muons or calorimeter 
clusters to identify them and measure their momenta. The fringe field of the magnet 
between the VELO and TT is used to determine the momenta of particles with a 
resolution of 20-40\% and events are selected based on tracks with a large $p_T$ and
significant impact parameter to the primary vertex.
Finally the HLT will select the stored events with a frequency of 200 Hz.
The HLT algorithm has access to all data and starts by reconstructing 
the VELO tracks and the primary vertex. A fast pattern recognition program 
links the VELO tracks to the tracking stations T1-T3. The final selection of 
interesting events is a combination of the confirmation of the L1 decision with better 
resolution, and selection cuts dedicated to specific final states.
In addition to the events stored by HLT, 1.8 KHz will be stored to get systematics from
the data (e.g. a trigger on high di-muon mass will be used to calibrate tracking, 
inclusive \b $\to \mu$ events will be selected to calibrate trigger, and \Dstar events 
to calibrate the particle identification).
\section{LHCb expected physics performances}
\begin{figure}[h]
  \begin{center}
    \epsfig{figure=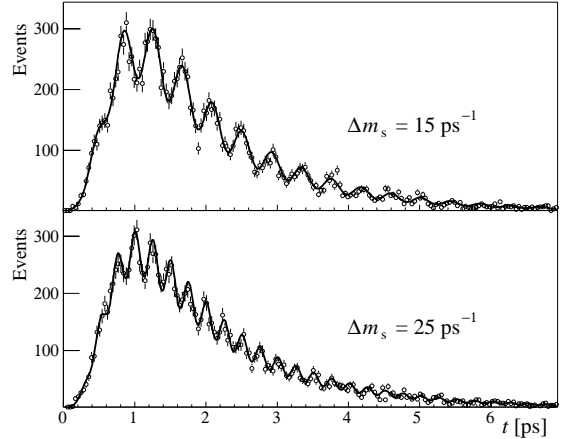,height=2.3in}
    \caption{\bsdsp decay rate for two different values of \deltams. Only \Bs decays 
      which have been tagged as not having oscillated are included. The curve shows 
      the result of the likelihood maximization.}
    \label{fig:dspi}
  \end{center}
\end{figure}

``Toy Monte Carlo'' programs have been used to estimate the LHCb sentivities to some 
\CP observables. These programs include signal 
resolution, efficiency, purity, etc. taken from studies with fully-simulated events.
Nevertheless, assumption need to be made concerning the properties of background events
due to the lack of statistics. In the real analyses, these properties and 
the systematics effects will be extracted from the data. Here only statistical errors 
are given. \newline
\indent The LHCb physics program includes many topics. For some of them, expected
sensitivities corresponding to one year of data taking 
(integrated luminosity of 2 fb$^{-1}$) and a \bbbar pair production cross-section 
assumed to be 0.5 mb will be given hereafter. 
\Bd mixing phase will be measured with \bpsiks decay with a sensitivity $\sim$0.02. 
\Bs mixing phase $\phi_s$ and decay-width $\Delta \Gamma_s/\Gamma_s$ will be 
assessed with \bsjpsiphi decay. This channel presents several 
challenges. Indeed, an angular analysis is needed as \jpsi and \mphi are vector mesons.
In addition, the oscillation frequency \deltams is expected to be large, requiring 
excellent proper-time resolution. The expected sensitivities to $\phi_s$ and 
$\Delta \Gamma_s/\Gamma_s$ are respectively 0.064 and 0.018.
\deltams will be measured with \bsdsp decay that is flavor-specific and give access 
to \deltams through flavour asymmetry analyses (see Fig. \ref{fig:dspi}). 
The sensitivity on \deltams is 0.009 (0.016) for \deltams = 15 ps$^{-1}$ (30 ps$^{-1}$).
Moreover, oscillations can be observed (5$\sigma$) for \deltams values up to 68 ps$^{-1}$.  
 \newline
\indent The measurement of the angle $\gamma$ of the unitarity triangle will be done in 
three different ways. Time-dependant decay asymmetries in \bsdsmpkpm decays combined 
with $\phi_s$ measurement from \bsjpsiphi decay, will give 
$\sigma(\gamma)=14^\circ-15^\circ$ without theoretical uncertainty. Time-dependant 
\CP asymmetries in \bdpippim and \bskpkm decays, in combination with measurement 
with \bpsiks and \bsjpsiphi respectively, will give $\sigma(\gamma)=4^\circ-6^\circ$
assuming U-spin symmetry \cite{fleisher}. Time-integrated rates of \bzdzkst, 
\bzdzbkstz, and \bzdzcpkstz decays \cite{gronau_wyler_dunietz}, will give 
$\sigma(\gamma)=7^\circ-8^\circ$. In the presence of new physics, these three 
different approaches could allow to identify it.
In addition, very rare decays like \bzsmumu decay, \b $\to$ \s penguin processes 
as \bphiks, \bzdmumukst,\bzdkstargamma, \bzsphigamma and \bzsphiphi decays, 
\b $\to$ \d penguin processes with \bropi decay, \Bc mesons and \b-baryons will be 
studied in great detail by LHCb.

\end{document}